\documentclass[final,5p,times,twocolumn]{elsarticle}
\usepackage{graphicx}
\usepackage{amsmath}
\usepackage{amssymb}
\usepackage{bm}
\usepackage{mathrsfs}
\usepackage{color}
\usepackage{slashed}
\usepackage{dcolumn}

\newcommand{\bald}[1]{{\bf #1}}

\newcommand{\eqf}[1]{\begin{equation}\begin{split}#1\end{split}\end{equation}}

\journal{Nuclear Physics B}

\begin{document}

\begin{frontmatter}



\title{Parton showers as sources of energy-momentum deposition in the QGP and their implication
for shockwave formation at RHIC and at the LHC}



\author[]{R. B. Neufeld}
 \address[]{Los Alamos National Laboratory, Theoretical Division, MS B238, Los Alamos, NM 87545, U.S.A.}
 \ead{neufeld@lanl.gov}
 \author[]{Ivan Vitev}
   \ead{ivitev@lanl.gov}

\begin{abstract}
We derive the distribution of energy and momentum transmitted from a primary fast parton and its 
medium-induced bremsstrahlung gluons to a thermalized quark-gluon plasma. Our calculation takes into 
account the  important and thus far neglected effects of quantum interference between the resulting 
color currents. We use  our  result to obtain the rate at which energy is absorbed by the medium as 
a function of time and find that  the rate is modified by the quantum interference between 
the primary parton and secondary gluons. 
This Landau-Pomeranchuk-Migdal type interference persists for time scales relevant to heavy ion 
phenomenology.   We further couple the newly derived source of energy and momentum deposition to 
linearized hydrodynamics to obtain the bulk medium 
response to realistic parton propagation and splitting in the quark-gluon plasma. We find that 
because of  the characteristic large angle in-medium  gluon emission and the multiple sources of energy 
deposition in a parton shower, formation of well defined Mach cones by energetic jets in heavy 
ion reactions is not likely.
\end{abstract}




\end{frontmatter}


\section{Introduction}
\label{introduction}

The suppression in the production rates of energetic leading particles and jets in relativistic heavy 
ion reactions relative to a naive 
superposition of nucleon-nucleon collisions is one of the most striking results from the heavy ion program 
at the Relativistic Heavy Ion Collider (RHIC)~\cite{Adcox:2001jp} and now at the Large Hadron Collider 
(LHC)~\cite{LHC,LHCII}.  These phenomena, often collectively referred to as ``jet quenching''~\cite{Gyulassy:2003mc}, 
have been studied extensively both experimentally and theoretically  and are thought to provide valuable 
information about the  quark-gluon plasma (QGP) created in these events~\cite{qgpform}.

At a fundamental level, the physics of jet quenching is largely reflective of the interaction of energetic, 
or fast, partons with the QGP medium which they traverse. Fast partons lose energy primarily through medium-induced 
radiation~\cite{jet1}.  Thus, the original fast parton evolves into an in-medium parton shower. The shape 
and energy distribution associated with this shower provide more information about the underlying QCD 
dynamics than the suppression 
of leading particles alone and form the basis for using full jet observables as a new and powerful probe 
of the QGP~\cite{jetty}.

Parton showers can also be substantially modified through collisional energy losses to the underlying medium.  
Although the collisional energy loss associated with a single parton is thought to be relatively small, 
the cumulative effect associated with a full parton shower can become quite large as the radiated gluons 
themselves become sources of collisional energy loss.  Thus, the question of how much energy a parton 
shower transmits to the medium is of fundamental interest in the description of full jet observables.

A closey related question is how the medium responds to a parton shower. As a parton shower loses energy and 
momentum, the underlying medium is affected in a way that depends on its properties  as well 
as the space-time distribution of the energy and momentum deposition.  This problem has gained attention 
due to experimental measurements of azimuthal particle correlations associated with high $p_T$ triggers 
in heavy ion collisions that display a double-peaked or conical structure~\cite{machexp}.  These 
measurements may reflect the interaction of fast partons with the medium and among the proposed explanations 
for the structure are Mach cone shockwaves excited in the bulk medium by fast partons~\cite{machprop}.  
Other explanations that do not reflect the interaction of fast partons with the medium, such as 
fluctuating initial conditions and triangular flow have also been proposed~\cite{Alver:2010gr}.  Whether 
or not the conical structure associated with high $p_T$ triggers at RHIC is associated with the medium 
response to a parton shower, this topic remains important, particularly in the light of the 
the new exciting possibilities for jet physics with the LHC heavy ion program.

Both the question of how much energy a parton shower transmits to the medium and how the medium responds 
to a parton shower can be addressed in the same framework by calculating the source term associated 
with the parton shower. The source term, here denoted $J^\nu(x)$, is the space-time distribution of 
energy and momentum flowing between a parton shower and the underlying medium. It couples to the 
energy-momentum tensor (EMT)  
$T^{\mu\nu}$ as $\partial_\mu T^{\mu\nu} = J^\nu$.  The source term carries information about the rate 
of energy transfer to the medium and acts as a source (hence the name ``source term'') for the evolution 
of the underlying medium in the presence of the parton shower.

Previous calculations of the source term have focused on a single primary parton~\cite{Neufeld:2008hs}.  
Applications to shockwave formation that have attempted to include medium-induced gluon bremsstrahlung 
have neglected the spatial extent of the source and, just as importantly, the color quantum interference 
between the primary parton and associated radiation~\cite{Neufeld:2009ep}.
In what follows we will perform the first calculation of the source term associated with a parton shower.  
Our calculation will include the full spatial extent and color quantum 
interference effects, and will be presented in closed analytic form.  To perform this calculation, we will 
start with the result of a recent paper by Neufeld ~\cite{Neufeld:2010xi} in which the source term induced by 
a single fast parton in a medium of thermal quarks was derived in integral form using thermal field theory.  
The source term was obtained 
by taking the thermal average of the divergence of the quark energy-momentum tensor, with the fast parton 
coupled to the medium by adding an interaction term to the Lagrangian.

Our paper is organized as follows: we present the theoretical formalism for the evaluation of the source 
term associated with a parton shower in section~\ref{formalism}. In section~\ref{linearhydro} we present 
two applications of the source term we derive: the rate of energy transfer between the parton shower and 
the medium and the response of the medium to the parton shower.  We show that a significant amount of 
energy can be transferred from the parton shower to the medium depending sensitively on the average angle 
of gluon radiation.  We further show that for realistic average angle of gluon emission the medium response 
no longer appears as a well defined Mach cone but rather a superposition of several distinct perturbations 
in the medium.  Our summary and conclusions are given in section~\ref{conclude}.

\section{Theoretical formalism}
\label{formalism}

Our approach in this section is to extend the result presented in~\cite{Neufeld:2010xi} to the case of 
an asymptotically 
propagating primary fast parton which undergoes medium-induced  gluon bremsstrahlung.  In this manuscript 
we use the following 4-vector notation: $x = (t,\bald{r}), u = (1,\bald{u})$.
\begin{figure*}
\centerline{
\includegraphics[width = 0.32\linewidth]{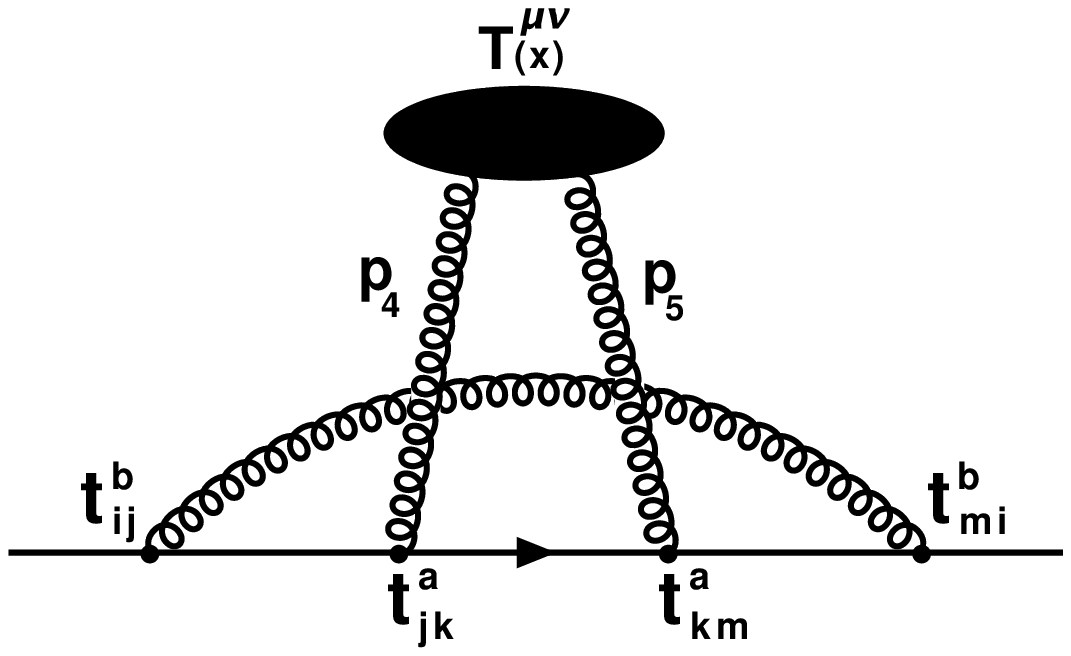}\hskip0.01\linewidth
\includegraphics[width = 0.32\linewidth]{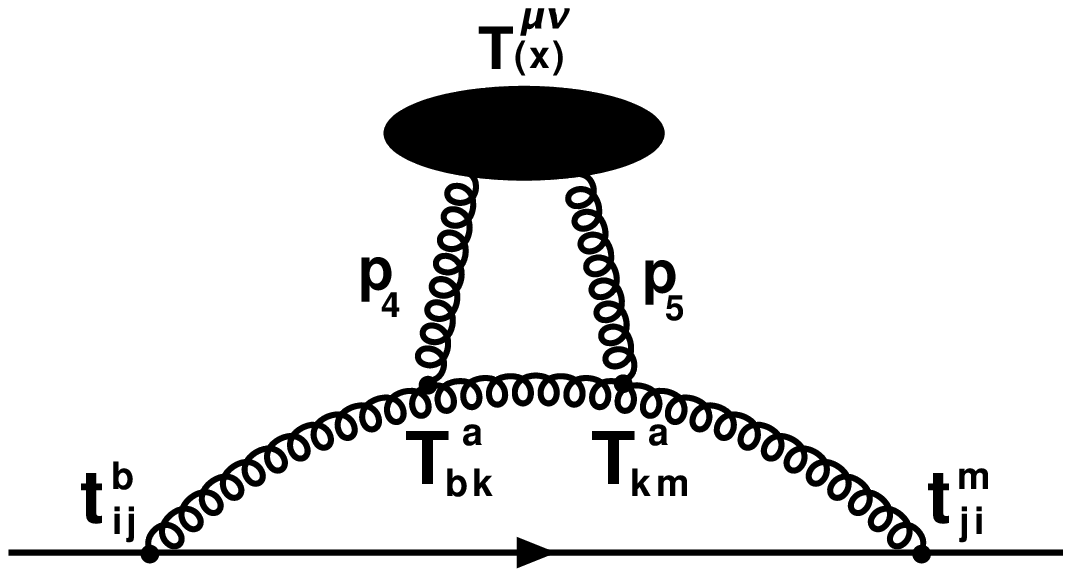}\hskip0.01\linewidth
\includegraphics[width = 0.32\linewidth]{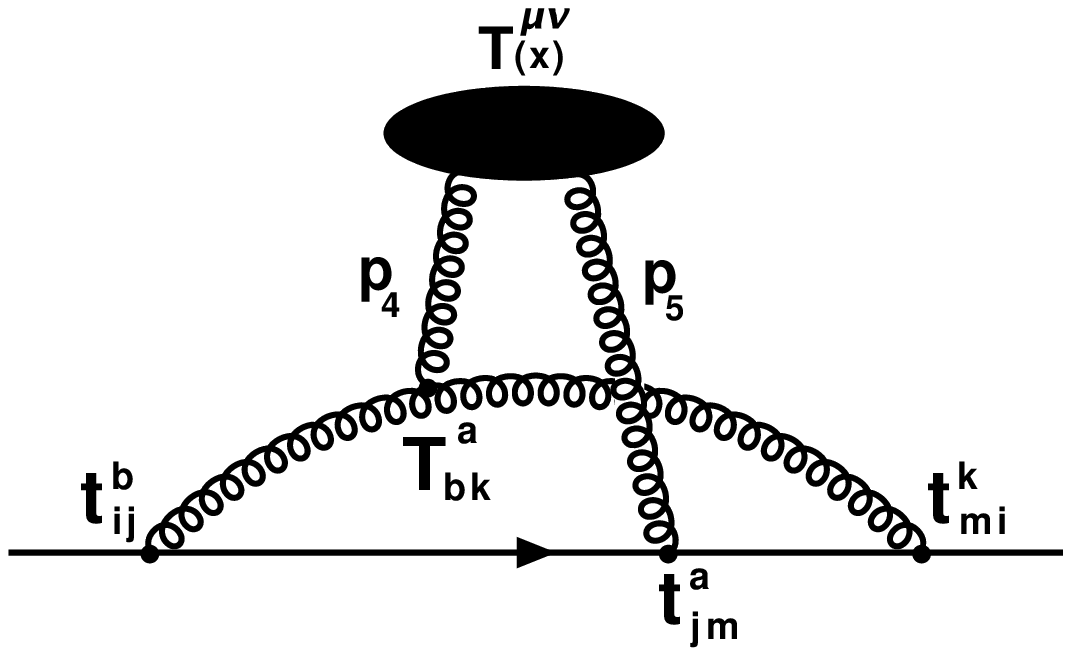}\hskip0.01\linewidth
}
\caption{Feynman diagrams for the source term  induced by a primary parton that undergoes 
medium-induced bremsstrahlung.  The color structure of the different contributions to the 
source term is explicitly shown. The third diagram represents the non-trivial interference 
between the parent parton and the radiated gluon (the second way of attaching $p_4$ and $p_5$ 
to the source is not shown explicitly).
}
\label{feyn1}
\end{figure*}

Our starting points are the Feynman diagrams of Figure~\ref{feyn1}.  The dark blobs represent 
the divergence of the medium energy-momentum tensor, for which we will implement the 
result derived in~\cite{Neufeld:2010xi}. In the left diagram of Figure~\ref{feyn1} the gluon 
lines labeled with $p_4$ and $p_5$ connect the medium to the primary fast parton (a quark in the diagram), 
in the middle diagram they connect to the radiated gluon, and in the right diagram they connect to 
both. This diagram represents the non-trivial quantum interference between the two partons. A fourth 
diagram  with $p_4$ and $p_5$ exchanged in the interference term is not shown in Figure~\ref{feyn1} but is 
included in the calculation presented here.

In Figure~\ref{feyn1} the {\it SU(3)} color matrices are denoted by upper and lower case $T$ and $t$ 
for the adjoint and fundamental representations, respectively.  Using standard color algebra,
we find that the color factors for the first two diagrams in Figure~\ref{feyn1} simply yield the 
quadratic Casimir of the particular representation: $C_F=4/3$ for a quark (first diagram) and $C_A=3$ 
for a gluon (middle diagram).  The color factors for the third and fourth (not shown)  diagrams yield  
$-C_A/2 = -3/2$ each.  We point out that the color factors of these last two contributions remain the same 
if the primary parton is a gluon. Note that we have averaged over the initial colors and factored out the
Casimir associated with the bremsstrahlung vertex.

If for the moment we write the divergence of the medium EMT as an arbitrary function $J^\nu(x,u_1,u_2)$ 
(with $u_1$ and $u_2$ being the velocity 4-vectors of the primary parton and radiated gluon, respectively), 
then the combined result for the four contributing diagrams is: 
\begin{eqnarray}
\label{basicform}
\partial_{\mu} T^{\mu\nu} &=&C_p J_a^\nu(x,u_1,u_1) + C_A J^\nu(x,u_2,u_2) \nonumber \\ 
   && - \frac{C_A}{2}\left[J^\nu(x,u_1,u_2) + J^\nu(x,u_2,u_1)\right]\, , 
\end{eqnarray}
where $C_p$ is the quadratic Casimir of the primary parton and the subscript $a$ indicates that 
we consider the primary parton to propagate asymptotically.  In the limit of perfectly collinear 
radiation one has $u_1 = u_2$ and the interference terms exactly cancel the contribution from the 
radiated gluon.  This result reflects the fact that the  medium will never resolve the bremsstrahlung  
gluon from the parent parton and the color of the system is the 
same before and after the radiation.  However, if  radiation is not perfectly collinear 
(and $u_1 \neq u_2$), the system evolves in time, the two new color currents separate,  
and the interference terms gradually vanish. At $t\rightarrow +\infty$ the system begins to look 
like two separate color charges.

In this work we consider the primary parton to be propagating asymptotically, that is we ignore any finite 
time effects associated with the initial large $Q^2$ scattering in a heavy ion collision. This problem has been examined 
in a different context in~\cite{finitetime}.  Although these effects are important to understand and may 
very well be significant, our focus is instead on the time dependence associated with the gluon radiation 
and the interference effects that arise with it.  When both gluon lines 
attach to the primary quark, the current can be represented by the classical asymptotic 
form: $j^\mu \rightarrow g Q_p^a\, u_1^\mu\,\delta(\bald{r} - \bald{u}_1\,t)$ with $Q^a_p\,Q^a_p = C_p$.  
Similarly, when both gluon lines attach to the secondary gluon the current can be represented 
by the form $j^\mu \rightarrow \Theta(t)\,g Q_g^a\, u_2^\mu\,\delta(\bald{r} - \bald{u}_2\,t)$.  
However, when the gluon lines are connected as in the third diagram of Figure~\ref{feyn1} 
there is no simple way to express the current and one must combine the product.  
The result is reflected in Eq.~(\ref{basicform}): $j^\mu(x_1)\,j^{\nu*}(x_2)\rightarrow 
-\Theta(t_1)\Theta(t_2) \,g^2 \frac{C_A}{2}u_1^\mu\delta(\bald{r}_1 
- \bald{u}_1\,t_1)u_2^\nu\delta(\bald{r}_2 - \bald{u}_2\,t_2)$, where we only consider 
interference effects after the point of emission.

With these considerations in place, we write the integral expression for the source term derived 
in~\cite{Neufeld:2010xi} for an asymptotically propagating fast parton in a medium of thermal quarks, 
corresponding to the first 
diagram of Figure~\ref{feyn1} and written in the notation of Eq.~(\ref{basicform}):
\begin{eqnarray}
\label{fullsourcetermA}
&&\hspace*{-1cm} J_a^\nu(x,u_1,u_1) = -4 i \,N_F\,g^4 \int \frac{d^4 p_3\,d^4 p_4\,d^4 p_5}{(2\pi)^{9}} 
e^{-i x \cdot(p_4 + p_5)} n_F(p_3) \nonumber  \\
&&\hspace*{-.2cm}\times \, \delta(p_3^2)G_R(p_4)G_R(p_3 + p_4)G_R(p_5)
\delta(p_4\cdot u_1)\delta(p_5\cdot u_1) \nonumber \\
&& \hspace*{-.2cm}\times\, \left[(2(p_3\cdot u_1)^2 - u_1^2 p_3\cdot p_4)p_5^\nu \right.  \nonumber \\
&&\hspace*{.2cm} \left. - u_1^\nu(p_3\cdot u_1)(2 p_3\cdot p_5 + p_4\cdot p_5)\right] \, , 
\end{eqnarray}
where $G_{R}(p) = (p^2 + i\epsilon p^0)^{-1}$ is the retarded Green's function, $n_F(p) = (e^{|p^0|/T} + 1)^{-1}$ 
is the Fermi distribution function, $T$ is the temperature, and $N_F$ is the number of active flavors.  

For our calculation we also need the generalization of Eq.~(\ref{fullsourcetermA}) to the case of two 
separate velocities, that is $J^\nu(x,u_1,u_2)$, created at an initial time $t = 0$.  The primary 
effect of this modification enters through the $\delta$ function structure of Eq.~(\ref{fullsourcetermA}): 
\eqf{\label{fundreplace}
\delta(p_4\cdot u_1)\delta(p_5\cdot u_1) \rightarrow 
\frac{-1}{(2\pi)^2(p_4\cdot u_2 + i\epsilon)(p_5\cdot u_1 + i\epsilon)}\, .
} 
The above replacement can easily be understood from the integral representation of the $\delta$ function: 
$2\pi \delta(p\cdot u) = \int_{-\infty}^\infty d t\, e^{i t p\cdot u}$.  For the case of a charge 
created at time $t = 0$ the integration begins at 0 rather than $-\infty$.  One can see 
that after contour integration in the $p_4^0,p_5^0$ variables the primary result of the replacement 
made in Eq.~(\ref{fundreplace}) is that $p_4$ and $p_5$ no longer enter the exponential symmetrically when 
$u_1 \neq u_2$.  The result is that the contribution from $J^\nu(x,u_1,u_2)$ will die out on an timescale related to 
the inverse of the difference between $u_1$ and $u_2$.

The specific generalization of the polynomial structure in the numerator (the third and fourth lines) of 
Eq.~(\ref{fullsourcetermA}) to the case of two separate velocities is not as essential as the replacement 
shown in Eq.~(\ref{fundreplace}) (or the pole structure). In particular, corrections to 
the form of Eq.~(\ref{fullsourcetermA}) are 
at most proportional to $\bald{u}_1 - \bald{u}_2$, which should be suppressed when the radiation angle is 
smaller than one.  
With this in mind, we make the obvious assignment that the first term remain symmetric in $u_1,u_2$.  
For the second term we use the form: 
\eqf{
u_1^\nu(p_3\cdot u_1)\delta(p_4\cdot u_1)\delta(p_5\cdot u_1)\rightarrow 
\frac{-u_1^\nu(p_3\cdot u_2)}{(2\pi)^2(p_4\cdot u_2 + i\epsilon)(p_5\cdot u_1 + i\epsilon)}\, ,
} 
based on the relation of $p_3$ and $p_4$ in the Green's function structure.  
In the worst case, the correction in this second term is proportional to $\bald{u}_1 - \bald{u}_2$.  
The combined result can be written as:
\begin{eqnarray}
\label{fullsourceterm}
&&  \hspace*{-1cm} J^\nu(x,u_1,u_2) = 4 i \,N_F\,g^4 \int \frac{d^4 p_3\,d^4 p_4\,d^4 p_5}{(2\pi)^{11}} 
e^{-i x \cdot(p_4 + p_5)} n_F(p_3)  \nonumber \\
&&  \hspace*{-.2cm} \times\, \frac{  \delta(p_3^2) G_R(p_4)G_R(p_3 + p_4)G_R(p_5)}
{(p_4\cdot u_2 + i\epsilon)(p_5\cdot u_1 + i\epsilon)} \nonumber \\
&&  \hspace*{-,2cm} \times \, \left[(2(p_3\cdot u_1)(p_3\cdot u_2) 
- u_1\cdot u_2 \,p_3\cdot p_4)p_5^\nu \right. \nonumber \\
&&  \hspace*{.2cm} \left. - u_1^\nu(p_3\cdot u_2)(2 p_3\cdot p_5 + p_4\cdot p_5) \right]\, .
\end{eqnarray}
The full source term, which is the sum of the terms in Eq.~(\ref{basicform}), can now be obtained by 
evaluating the expressions given in Eqs.~(\ref{fullsourcetermA}) and~(\ref{fullsourceterm}).  As we will show 
in what follows, one actually only needs to evaluate Eq.~(\ref{fullsourceterm}), for which 
Eq.~(\ref{fullsourcetermA}) can be obtained as a special case.  

We now proceed to evaluate Eq.~(\ref{fullsourceterm}).  In order to simplify the calculation we 
use bare propagators and  take the hard thermal loop (HTL) approximation.  
These simplifications will allow us to obtain an analytic result for the source term.  We also 
will consistently ignore radiative effects arising from the formation of a charge at an initial time. 
These effects are associated with causality and are in addition to the medium induced radiation we are 
considering in this paper. We postpone their consideration to a future work.  We emphasize 
that all vacuum contributions have already been subtracted from Eq.~(\ref{fullsourceterm}).  
Furthermore, although (\ref{fullsourceterm}) was derived for a medium of quarks (as can be seen  from 
the number of active flavors $N_F$ and the Fermi distribution $n_F(p)$)  the extension to 
include medium gluons is trivial in the HTL limit as will be shown below.

We obtain the HTL approximation by taking the limit that the fields generated by the 
external current are soft compared to the temperature, or formally $|p_4| \sim g T \ll T$ in 
Eq.~(\ref{fullsourceterm}). The coupling constant $g$ is formally considered to be much less than one.  
Explicitly, we expand $G_R(p_3 + p_4)$ in the limit of $|p_4| \ll |p_3|$ since $|p_3|$, which appears in 
the Fermi distribution function, is cut off in the integration at a value on the order of $T$ to 
lowest contributing order.  We find that in the HTL approximation Eq.~(\ref{fullsourceterm}) reduces to:
\begin{eqnarray}
\label{htlsourceterm}
 \!\!\!\!\!\!\!\!\! \!\!  \!\! J^\nu(x,u_1,u_2)_{\text{HTL}}\!\! \! &=&\!\!\!  -\frac{i\,m_D^2\,g^2}{2} \int 
\frac{d^4 p_4\,d^4 p_5}{(2\pi)^{8}}\int\frac{d\Omega}{4 \pi} e^{-i x \cdot(p_4 + p_5)} \nonumber \\
&&\times \frac{G_R(p_4)G_R(p_5)}{{(p_4\cdot u_2 + i\epsilon)(p_5\cdot u_1 + i\epsilon)}} \nonumber  \\
&&\times\left[\frac{p_4^2\left((v\cdot u_1)(v\cdot u_2)\,p_5^\nu 
- u_1^\nu\,(v\cdot u_2)(v\cdot p_5)\right)}{(p_4 \cdot v + i \epsilon )^2} \right.  \nonumber  \\
&& \left.+ \frac{u_1\cdot u_2 \,v\cdot p_4\,p_5^\nu + u_1^\nu\,
(v\cdot u_2)(p_4\cdot p_5)}{p_4 \cdot v + i \epsilon}\right]\, , 
\end{eqnarray}
where $m_D^2 = g^2\,T^2 \,N_F/6$ for a thermal medium comprised quarks only.  In Eq.~(\ref{htlsourceterm}) 
$d\Omega$ is the integration measure over the solid angle defined by the unit vector $\bald{v}$, 
and $v^\mu = (1,\bald{v})$.  As mentioned previously, the extension to include medium gluons is straightforward 
and is done with the modification $m_D^2 \rightarrow g^2\,T^2 (1 + N_F/6)$.

In order to evaluate Eq.~(\ref{htlsourceterm}) it is convenient to note that the $p_4$ and $p_5$ 
integrations separate.  The $p_5$ contribution has the same general form in each term which 
we evaluate as: 
\begin{eqnarray}
\label{greenie}
&& \!\!\!\!\!\!\!\!\!  \!\!\!\!\!\!\!\!\! 
\int \frac{d^4 p_5}{(2\pi)^{4}} \frac{\left(p_5^0,\bald{p}_5 \right)\, e^{- i x \cdot p_5}}{(p_5\cdot u_1+i\epsilon)}\,G_R(p_5) 
\nonumber \\
&&\!\!\!\!\!\!\!\!\!  \!\!\!\!\!\!\!\!\!  
 = - \Theta(t^2 - r^2) \frac{\gamma_1\left(\gamma_1^2(\bald{u}_1\cdot\bald{r}-u_1^2\,t),\bald{r} - \bald{u}_1\, 
\gamma_1^2(t - \bald{u}_1\cdot\bald{r})\right)}{4\pi(r^2 - t^2 + \gamma_1^2(t - \bald{u}_1 \cdot \bald{r})^2)^{3/2}} \, , 
\end{eqnarray}
where $\gamma_1 = 1/\sqrt{1-u_1^2}$, $t > 0$, and  we ignore radiative effects on the light cone associated 
with the formation of the field.  The result of Eq.~(\ref{greenie}) can be derived in a 
straightforward manner by expressing components of $p_5$ as a derivative on $x$ and using the 
relationship that $\int d^4 p e^{- i x \cdot p} G_R(p) = (2\pi)^3\delta(t^2 - r^2)$ along 
with $(p_5\cdot u_1+i\epsilon)^{-1} = - i \int_0^{\infty} d w \,e^{i \,w p_5\cdot u_1}$.  
The step function in Eq.~(\ref{greenie}) is a result of causality and ensures that the final result 
for the source term will only contribute in the time-like region.

The $p_4$ integration in Eq.~(\ref{htlsourceterm}) cannot be separated from the $\bald{v}$ 
integration and thus is more involved.  For the first term one is aided by the cancellation occurring 
from $p_4^2 \,G_R(p_4) = 1$.  Again using $(p_4\cdot u_2+i\epsilon)^{-1} = - i \int_0^{\infty} d w \,e^{i \,w p_4\cdot u_2}$ 
and writing $(p_4 \cdot v + i \epsilon )^{-2} = -d/dp_4^0 (p_4 \cdot v + i \epsilon )^{-1}$ followed by 
an integration by parts allows:
\begin{eqnarray}
&& \!\!\!\!\!\!\!\!\!     \!\!\!\!\!\!\!\!\!     \int \frac{d^4 p_4}{(2\pi)^{4}} \int\frac{d\Omega}{4 \pi} 
e^{-i x \cdot p_4} \frac{1}{{(p_4\cdot u_2 + i\epsilon)}} \frac{g(v)}{(p_4 \cdot v + i \epsilon )^2}  \nonumber \\
&& \!\!\!\!\!\!\!\!\!     \!\!\!\!\!\!\!\!\!  =   i \int_0^\infty  d w \, d\tau \int \frac{d^4 p_4}{(2\pi)^{4}} 
\int\frac{d\Omega}{4 \pi}e^{-i p_4\cdot(x - u_2 w - v \tau)}(t-w)\,g(v) \, ,
\end{eqnarray}
where $g(v) = \left((v\cdot u_1)(v\cdot u_2)\,p_5^\nu - u_1^\nu\,(v\cdot u_2)(v\cdot p_5)\right)$.  
At this point the $p_4$ integration is performed to yield a $\delta$ function, which enables the 
remaining integrals to be performed analytically.  The result is obtained as
\begin{eqnarray}
&&\int \frac{d^4 p_4}{(2\pi)^{4}} \int\frac{d\Omega}{4 \pi} e^{-i x \cdot p_4} 
\frac{1}{{(p_4\cdot u_2 + i\epsilon)}} \frac{g(v)}{(p_4 \cdot v + i \epsilon )^2}  \nonumber \\
&&= \frac{i\gamma_2\,g(v = v')\Theta(t^2 - r^2) }{4\pi \sqrt{r^2 - t^2 
+ \gamma_2^2(t - \bald{u}_2\cdot \bald{r})^2}} \, ,
\end{eqnarray}
where
\eqf{
v' &= \left(1,\frac{\bald{r} - \bald{u}_2 t}{\left(\gamma_2^2 
(\bald{u}_2\cdot \bald{r} - u_2^2\,t) +\gamma_2 \sqrt{r^2 - t^2 + 
\gamma_2^2(t - \bald{u}_2\cdot \bald{r})^2} \right)} + \bald{u}_2\right).
}

Lastly, we must evaluate:
\eqf{\label{last}
\int \frac{d^4 p_4}{(2\pi)^{8}}\int\frac{d\Omega}{4 \pi}\frac{e^{-i x \cdot p_4} 
G_R(p_4)}{{(p_4\cdot u_2 + i\epsilon)}}\left[u_1\cdot u_2 p_5^\nu + \frac{u_1^\nu 
(v\cdot u_2)(p_4\cdot p_5)}{p_4 \cdot v + i \epsilon}\right].
}
The first term in Eq.~(\ref{last}) can be obtained in the same manner as Eq.~(\ref{greenie}) with the result: 
\eqf{\label{greenieII}
\int &\frac{d^4 p_4}{(2\pi)^{4}} \frac{e^{- i x \cdot p_4} G_R(p_4) }{(p_4\cdot u_2+i\epsilon)}= \frac{i \gamma_2 \, 
\Theta(t^2 - r^2) }{4\pi\sqrt{r^2 - t^2 + \gamma_2^2(t - \bald{u}_2\cdot \bald{r})^2}}.
}
The second term is more involved and for the sake of brevity we briefly outline the important steps in obtaining the result.  
We first write $(p_4 \cdot v + i\epsilon)^{-1} = - i \int_0^{\infty} d \tau \,e^{i \,\tau p_4\cdot v}$ and then use
 contour methods to perform the $p_4^0$ integration (which places a restriction that $\tau <  t$).  One can next 
perform the $\tau$ integration to leave: 
\begin{eqnarray}\label{midform}
&& \!\!\!\!\!\!\!\!\!     \!\!\!\!\!\!\!\!\! \int \frac{d^4 p_4}{(2\pi)^{4}}\int\frac{d\Omega}{4 \pi}
\frac{e^{-i x \cdot p_4} G_R(p_4)}{{(p_4\cdot u_2 + i\epsilon)}}
\left[\frac{(v\cdot u_2)p_4^\alpha}{p_4 \cdot v + i \epsilon}\right]   \nonumber  \\
&& \!\!\!\!\!\!\!\!\!     \!\!\!\!\!\!\!\!\!  = - i \int \frac{d p\, d\Omega_1}{(2\pi)^3}\frac{d\Omega}{4\pi}
\left[ \text{Im} \frac{i v_1^\alpha e^{-i p (x \cdot v_1)}}{(u_2\cdot v_1)}  
\frac{(1-e^{i t p(v_1\cdot v)})}{(v_1\cdot v)}(u_2\cdot v) \right.  \nonumber \\
&& \!\!\!\!\!\!\!\!\!     \!\!\!\!\!\!\!\!\! \left.+ (u_2\cdot v) \frac{1 - e^{i t p \bald{v}_1 \cdot 
(\bald{u}_2 - \bald{v})}}{\bald{v}_1 \cdot (\bald{u}_2 - \bald{v})}
e^{i p \,\bald{v}_1\cdot(\bald{x} - \bald{u}_2 t)} \left[ \frac{\left(\bald{u}_2\cdot\bald{v}_1,\bald{v}_1\right)^\alpha}
{(\bald{u}_2\cdot\bald{v}_1)^2 - 1}\right]\right].
\end{eqnarray}

The second line of Eq.~(\ref{midform}) can be simplified by redefining the $v$ angular variable in terms of 
$\bald{v}_1\cdot \bald{v}$: that is, let $\bald{v}_1\cdot \bald{v} \equiv \cos\theta$, where $\theta$ is 
associated with $d\Omega$.  One can further use the relations $\int d \Omega \, e^{i \bald{k}\cdot \bald{r}} 
= 4 \pi \sin k r/k r$ and $\int d \Omega \,\bald{k}  \, e^{i \bald{k}\cdot\bald{r}} = 4 \pi \.i \,\bald{r} 
(\sin k r - k r \cos k r)/k \, r^3$ to reduce the expression to two integrals which can be performed in closed 
form using standard integral relations.  The result is:
\begin{eqnarray}
&&  \!\!\!\!\!\!\!\!\!     \!\!\!\!\!\!\!\!\!  - i \int \frac{d p\, d\Omega_1}{(2\pi)^3}\frac{d\Omega}{4\pi} \text{Im} 
\frac{i (1,{\bald v}_1) e^{-i p (x \cdot v_1)}}{(u_2\cdot v_1)}  \frac{(1-e^{i t p(v_1\cdot v)})}{(v_1\cdot v)}(u_2\cdot v) \nonumber \\
&& \!\!\!\!\!\!\!\!\!     \!\!\!\!\!\!\!\!\!  = \frac{-i}{4\pi r}\left(\left(\frac{r-t \tanh^{-1}[\frac{r}{t}]}{t}\right)\, ,
\frac{\bald{r}}{r}\left(\frac{r-t \tanh^{-1}[\frac{r}{t}]}{r}\right)\right)  \, , 
\end{eqnarray}
where for simplicity we have restricted our attention to the region of $r < t$ knowing from Eq.~(\ref{greenie}) that this is 
the only region that contributes in the final expression.

Finally we consider the third line of Eq.~(\ref{midform}), which can again be simplified by redefining the $v$ angular 
variable in terms of $\bald{v}_1\cdot \bald{v}$.  One can further use the integral relations 
$\int_0^{2 \pi} d \phi [1,\cos\phi]\exp {i k b \cos [\phi-\alpha]} = 2\pi[J_0(k b),i \cos\alpha J_1(k b)]$ and \cite{magnus}:
\begin{eqnarray}
\int_0^\infty d k \,e^{i k y}J_0(k b) = \frac{1}{\sqrt{b^2 - (y + i\epsilon)^2}}  \, , \\
\int_0^\infty d k \,e^{i k y}J_1(k b) = \frac{1}{b}\left(1 + \frac{i y}{\sqrt{b^2 - (y + i\epsilon)^2}}\right) \, , 
\end{eqnarray}
to again reduce the expression to two integrals.  In the above expressions $J_i$ are Bessel functions of the first kind, $b > 0$ and $\epsilon$ is a positive infinitesimal quantity.  Performing the final two integrations,  we find
\begin{eqnarray}
\label{involved}
&& \!\!\!\!\!\!\!\!\!  \!\!\!\!\!\!\!\!\!  \!\!\!\!\! - i \int \frac{d p\, d\Omega_1}{(2\pi)^3}\frac{d\Omega}{4\pi}
 (u_2\cdot v) \frac{1 - e^{i t p \bald{v}_1 \cdot (\bald{u}_2 - \bald{v})}}{\bald{v}_1 \cdot 
(\bald{u}_2 - \bald{v})}e^{i p \,\bald{v}_1\cdot(\bald{x} - \bald{u}_2 t)} 
\left[ \frac{\left(\bald{u}_2\cdot\bald{v}_1,\bald{v}_1\right)}{(\bald{u}_2\cdot\bald{v}_1)^2 - 1}\right] \nonumber \\
&& \!\!\!\!\!\!\!\!\!  \!\!\!\!\!\!\!\!\! \!\!\!\!\!  = i\frac{\gamma_2\left(u_2^2\,r^2 - (\bald{u}_2\cdot\bald{r})^2,  \bald{u}_2(r^2 - \bald{u}_2\cdot\bald{r}\, t) - \bald{r}(\bald{u}_2\cdot\bald{r} - u_2^2 t)\right)}
{4 \pi\,(\bald{r} - \bald{u}_2 t)^2\,\sqrt{r^2 - t^2 + \gamma_2^2(t - \bald{u}_2\cdot\bald{r})^2 }} \nonumber  \\
&& \!\!\!\!\!\!\!\!\!  \!\!\!\!\!\!\!\!\!  \!\!\!\!\! - i \frac{\tanh^{-1} u_2 - u_2}{4\pi \,t u_2}
\left(1,\frac{\bald{u}_2}{u_2^2}\right) 
+i \left(\frac{\bald{u}_2\cdot\bald{r} - u_2^2 t }{4 \pi(\bald{r} - \bald{u}_2 t)^2},
\frac{\bald{u}_2 t}{4\pi(\bald{r} - \bald{u}_2 t)^2}\right)\, .
\end{eqnarray}

We now have all of the necessary pieces to yield a final result for Eq.~(\ref{htlsourceterm}).  In order to achieve a 
compact form, is it helpful to define the notation $w_i^\alpha = x^\alpha - \gamma_i^2(x\cdot u_i)u_i^\alpha$, in which case one finds: 
\eqf{\label{primaryres}
&J^\nu(x,u_1,u_2)_{\text{HTL}} \\
& = \Theta(t^2 - r^2)\frac{m_D^2\,\alpha_s}{8\pi} \left[-\frac{\gamma\, w_1^\nu \left(\delta u \cdot w_2 
+ u_1\cdot u_2 \gamma_2 \sqrt{-w_2^2}\right)}{\left(w_2^0 +\gamma_2 \sqrt{-w_2^2} \right)^2(-w_1^2)^{3/2}} \right. \\
&+\frac{u_1^\nu \gamma_1 \left(\delta u \cdot w_1 \left(t - w_2^0 -\gamma_2 \sqrt{-w_2^2} \right) + w_1^2\right)}
{\left(w_2^0 +\gamma_2\sqrt{-w_2^2}  \right)^2(-w_1^2)^{3/2}} 
- \frac{u_1\cdot u_2 \gamma_2 \gamma_1 w_1^\nu }{\sqrt{-w_2^2}(-w_1^2)^{3/2}} \\
&+ \frac{u_1^\nu \gamma_1}{(-w_1^2)^{3/2}}
\left(1 - \frac{\gamma_1^2(t-\bald{u}_1\cdot\bald{r}) }{(\bald{r} - \bald{u}_2 t)^2}\right.\\
&\left.  \times  \, \delta \bald{u}\cdot \left((\bald{r} - \bald{u}_2 t) + \frac{\gamma_2 \left(\bald{u}_2 
(r^2 - \bald{u}_2\cdot\bald{r} t) + \bald{r}(u_2^2 t - \bald{u}_2\cdot\bald{r})\right)}{\sqrt{-w_2^2}} \right)\right) \\
&+ \frac{u_1^\nu \gamma_1}{(-w_1^2)^{3/2}}\left(\frac{u_2^2 t 
- \bald{u}_2\cdot\bald{r}+\delta\bald{u}\cdot \bald{u}_2\gamma_1^2(t-\bald{u}_1\cdot\bald{r})}{t}
\left(\frac{\tanh^{-1} u_2 - u_2}{u_2^3}\right) \right.\\
&\left.\left. + \frac{r-t \tanh^{-1}\frac{r}{t}}{r}\left(\frac{\gamma_1^2(\bald{u}_1\cdot\bald{r}-u_1^2\,t )}{t}-1 
+ \frac{(\bald{r}\cdot\bald{u}_1) \gamma_1^2(t - \bald{u}_1\cdot\bald{r})}{r^2}\right)\right) \right]\, , 
}
where we have defined $\delta u \equiv u_1 - u_2$.  Equation~(\ref{primaryres}) is our primary result and we now discuss 
its general features and also how to specialize it to the asymptotic case $J_a^\nu(x,u_1,u_1)$.  
Eq.~(\ref{primaryres}) is restricted by causality, which is contained in the step function $\Theta(t^2 - r^2)$. 
Furthermore, the first four lines of Eq.~(\ref{primaryres}) contribute in the asymptotic limit, whereas the last two lines 
are transient contributions, that is, they vanish as $t\rightarrow \infty$ even for $u_1 = u_2$.  This can be seen by 
taking the case of $u_1 = u_2$ and allowing $\bald{u}\cdot \bald{r} - u^2 t = {\rm const.}$, while taking $t\rightarrow \infty$.  
Thus one can obtain the asymptotic case $J_a^\nu(x,u_1,u_1)$ by removing the step function and the final two lines of 
Eq.~(\ref{primaryres}), and of course setting $u_1 = u_2$.  In this asymptotic limit our result should reduce to the 
result obtained previously by Neufeld in~\cite{Neufeld:2008hs}. We have verified numerically that this is indeed true.

\begin{figure}
\centerline{
\includegraphics[width = 0.95\linewidth]{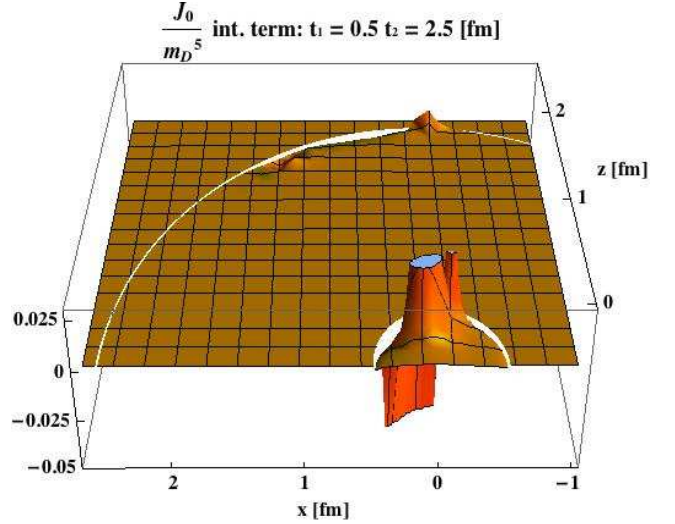}
}
\caption{A 3D representation of the extent and magnitude of the appropriately scaled interference term $J_0/m_D^5$ 
at two different times $t=0.5$~fm and $t=2.5$~fm for a primary parton that propagates in the $z$ direction. 
Note that this term disappears in the long time limit.  
}
\label{sourceplot}
\end{figure}

Recall from Eq.~(\ref{basicform}) that the total source term is given by the combination $C_p J_a^\nu(x,u_1,u_1) 
+ 3 J^\nu(x,u_2,u_2) - \frac{3}{2}[J^\nu(x,u_1,u_2) + J^\nu(x,u_2,u_1)]$, where the last two terms result from the 
interference graphs of Figure~\ref{feyn1} and we refer to this contribution as ``interference term'' in what follows.  
To help illustrate the result of Eq.~(\ref{primaryres}) we plot the interference term contribution for two different 
times in Figure~\ref{sourceplot} above.  For all numerical results we use the following parameters: $g = 2$, 
$T = 0.35$ GeV and $m_D = g T$, which are based on the average values obtained for LHC collision energies in a 
Bjorken expanding plasma and we take $N_F = 0$ for a gluon-dominated medium.  Additionally we impose medium-induced 
Debye screening through an exponential decay factor, $e^{-m_D(\rho_1 + \rho_2)/2}$, where $\rho_1$ and $\rho_2$ are distances transverse to the 
axes of propagation defined by $u_1$ and $u_2$, respectively.  The velocity of the primary parton is chosen along 
the $z$ axis with a magnitude determined by the energy and mass of the parton: 
$E = \gamma m$ with $m = g T/\sqrt{6}$ for a quark and $m = gT/\sqrt{2}$ for a gluon.  The magnitude of the velocity 
of the radiated gluon is determined in the same way as the primary parton, and its direction determined by the 
angle of emission relative to the direction of the primary parton's propagation.  For simplicity we will take 
the emission to occur in the $x-z$ plane unless otherwise stated. Thus, all quantities are determined by specifying 
the energies of the primary parton and the radiated gluon, as well as the angle of emission.

\begin{figure*}
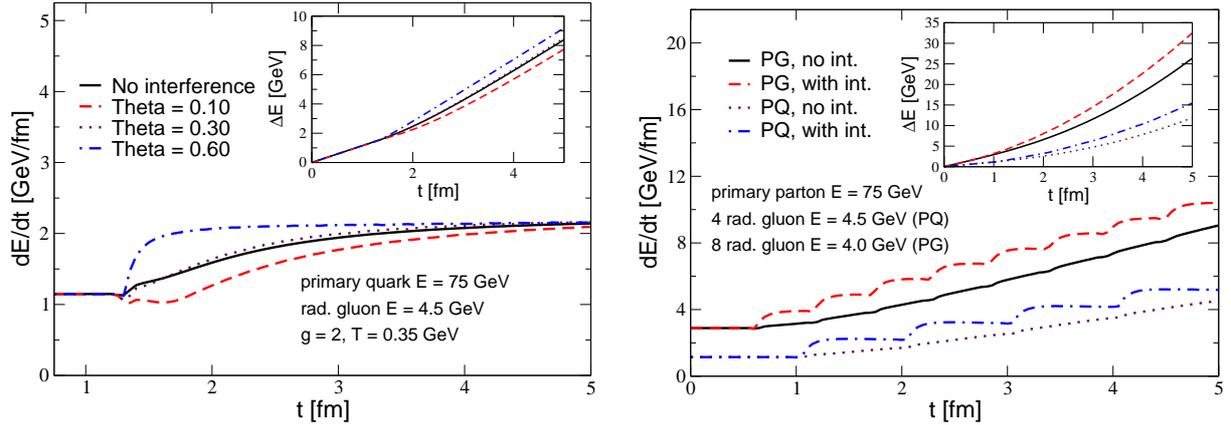

\centerline{
\includegraphics[width = 0.42\linewidth]{singlegluon_emission.eps}\hskip0.03\linewidth
\includegraphics[width = 0.42\linewidth]{multigluon_emission.eps}
}
\caption{Left panel: The rate of energy transfer from a $q\rightarrow q+g$ system to the QGP medium 
for different gluon emission angles $\theta = 0.1, \; 0.3, \; 0.6$. The naive no-interference result is
presented by a solid line. Insert shows the total energy absorbed by the medium as a function of time. 
Right panel: The same rate and net energy transfer are presented for realistic multiple gluon emission 
of average angle $\theta = 0.7$ for both $q \rightarrow q + 4\times g$  and $ g \rightarrow g + 8 \times g$ systems.}
\label{energyplot}
\end{figure*}

We plot the interference term contribution for two different times in Figure~\ref{sourceplot} above.  For this plot 
we choose a primary quark with energy $5$ Gev, a radiated gluon with energy $1$ GeV and angle of emission 
$\theta = 0.6$ (angles will always be given in radians).  Although these energies are phenomenologically small, 
they are convenient for the purposes of plotting.  The plot shows the zero component of the interference term 
contribution scaled to a dimensionless ratio for times $t = 0.5$~fm  and $t = 2.5$ fm (in the same figure), 
where time is measured from the moment of emission.  The plot shows that at early times when the two partons are 
close together their interference contribution is important, but at later times as they become separated the 
interference contribution diminishes and eventually vanishes.  In the next section we consider two applications 
of the source term derived in Eq.~(\ref{primaryres}).


\section{Applications: differential energy transfer to the medium and linearized hydrodynamics}
\label{linearhydro}

The main theoretical result of this paper is Eq.~(\ref{primaryres}), which combined with Eq.~(\ref{basicform}) 
gives the source term for a primary parton and a single radiated gluon as a function of time, including quantum 
color interference effects. In this section, we consider two applications of our result: the evaluation of the 
differential energy lost by a parton shower to the medium and the hydrodynamic response of the medium 
to a  parton shower.  

The differential energy transfer to the medium is obtained from the source term 
through a spatial integration of the zero component: 
\begin{eqnarray}
\frac{dE}{dt} &=& \int d\bald{r} \, \left[  C_p J_a^0(x,u_1,u_1) 
+ C_A J^0(x,u_2,u_2) \right. \nonumber \\
&& \left. - \frac{C_A}{2}(J^0(x,u_1,u_2) + J^0(x,u_2,u_1)) \right] \; . 
\label{etransfer}
\end{eqnarray}
This transfer is of a collisional nature and the terms involving only one velocity, $C_p J_a^0(x,u_1,u_1)$ 
and $C_A J^0(x,u_2,u_2)$, are fraught with an ultraviolet 
divergence typical of collisional energy loss~\cite{Thoma:1991ea}. We regulate these terms with a standard 
short distance cutoff $\Theta(\rho_i - 1/(2\sqrt{E_i T}))$ related to the finite momentum transfer, 
where $\rho_i$ is the distance transverse to the  axis of propagation defined by $u_i$ and $E_i$ is the 
corresponding energy.  On the other hand, the terms 
involving two velocities, $J^0(x,u_1,u_2)$ and $J^0(x,u_1,u_2)$, have no ultraviolet divergence and require 
no cutoff.  However, motivated by the finite kinematics and for consistency, we apply a short distance 
cutoff to these terms reflective of two distinct gluon exchanges: 
$\Theta(\rho_1 - 1/(\sqrt{E_1 T}))\Theta(\rho_2 - 1/(\sqrt{E_2 T}))$.  As these terms 
have no ultraviolet divergence, the sensitivity to the cutoff for the interference contribution is small.

\begin{figure*}
\centerline{
\includegraphics[width = 0.46\linewidth]{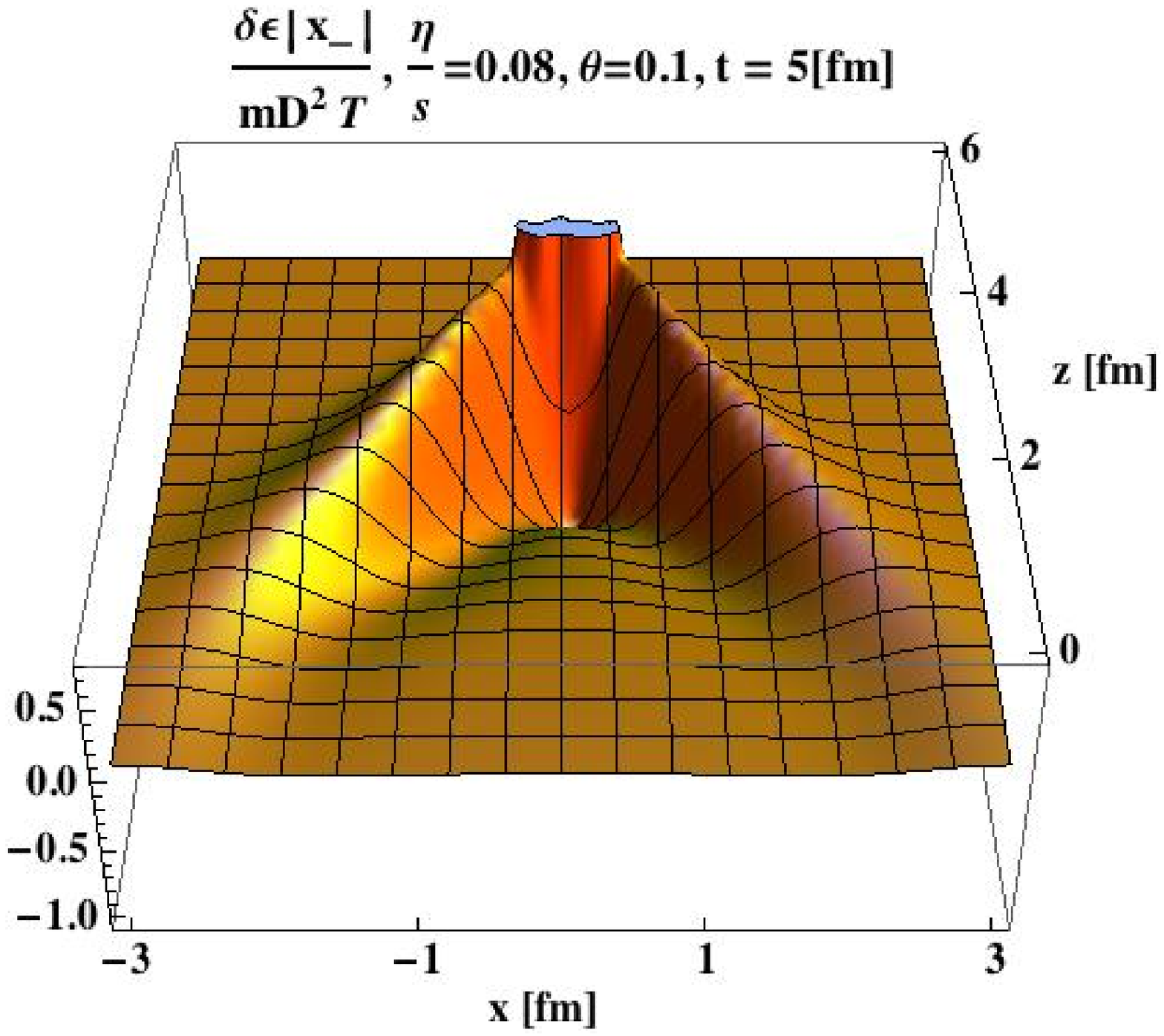}\hskip0.03\linewidth
\includegraphics[width = 0.46\linewidth]{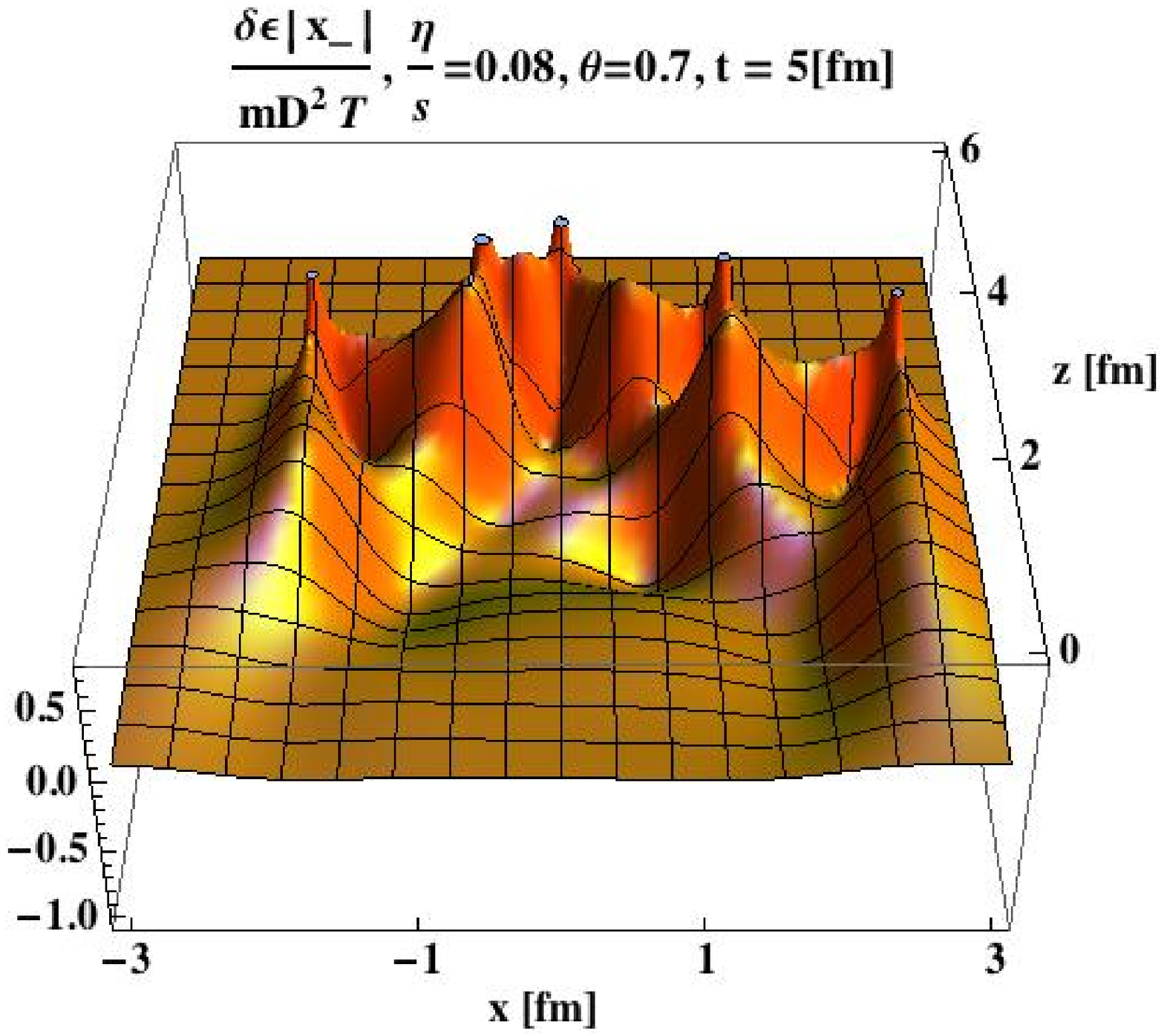}
}
\caption{In the left panel we show energy density disturbance generated by a parton shower originating from a 
primary quark with average gluon emission angle of $\theta = 0.1$, whereas in the right panel the average angle 
is $\theta = 0.7$.  One can see that for the narrow emission angle the parton shower generates a well defined 
Mach cone similar to what one would expect from a single primary parton.  However, the interference effects 
from such small angle emissions suppress the energy transfer to the medium.  For the larger emission angle there 
is no longer a well defined Mach cone but rather a superposition of several distinct perturbations in the medium.  
The interference effects for this angle actually lead to an enhancement in the overall energy gained by the medium, 
even relative to no interference.
}
\label{epsplots}
\end{figure*}

We now consider two scenarios for the differential energy transfer to the medium which are both presented in 
Figure~\ref{energyplot}.  The first is shown in the left panel and is for a primary quark of energy 
75~GeV which emits a gluon of energy 4.5 GeV at time $t = 1.25$~fm. At this point the primary 
quark's energy is reduced to 70.5 GeV. The differential energy loss to the QGP is shown for three different 
angles of emission, as well as without any interference effects. It should be noted that even in the absence of
interference it takes on the order of $\gamma/m_D$ for the radiative gluon to build up its asymptotic
energy transfer rate (see the solid curve). The strength of interference effects depends on the transverse 
separation of the two color currents. When these currents cannot be resolved by the medium the system behaves like
the parent parton. This is illustrated by the smallest emission angle we consider, $\theta = 0.1$. 
The red dashed curve shows the fact that interference suppresses the energy loss 
to the medium for times extending up to about 3-4 fm after emission.  For the larger angle of $\theta = 0.3$ there is a slight suppression 
initially followed by a slight enhancement persisting until about 2~fm after emission. Finally, we also consider 
$\theta = 0.6$ as we are motivated by the finding that the medium-induced bremsstrahlung is dominated by large 
angles~\cite{angle}.  For this largest angle the energy loss actually shows an enhancement relative to no 
interference which persists for about 2~fm after emission.  The insert shows the total energy transferred 
to the medium for the different angles as a function of time.

Next, we consider a more realistic representation of the medium-induced parton shower
inspired by the Gyulassy-Levai-Vitev (GLV) energy loss formalism~\cite{Vitev:2007ve}. Such parton showering 
is relevant to phenomenological applications that aim to elucidate the medium response to jet 
propagation in hot and dense QCD matter~\cite{machphenom}.  Specifically, we make use of the bremsstrahlung 
spectra for quark and gluon jets 
averaged over the collision geometry in central Pb+Pb reactions at the LHC that have been previously 
employed to discuss jet and particle production 
in heavy ion collisions at the highest $\sqrt{s_{NN}}$~\cite{lhcrad}. Medium-induced
gluon emission is probabilistic and depends on the parton species. The most likely numbers 
are 4 ($\omega_g \approx 4.5$~GeV)  and 8~($\omega_g \approx 4$~GeV)  for 75~GeV quark and gluon parent 
partons, respectively. We also note that the average emission angle is quite large, $\theta \approx 0.7$. 
The gluon emission points are uniformly distributed along the parton propagation path in a medium of 
length 5~fm. In the GLV approach the gluon distribution is azimuthally symmetric around the direction of
jet propagation. For simplicity, we distribute them here in the x-z plane.   Further, quantum interference effects 
are only taken between the primary parton and each radiated gluon - not between the multiple radiated gluons. 
The result is shown in the right panel of Figure~\ref{energyplot} where again we find an 
enhancement of energy transfer to the medium.  The insert shows the total energy transferred to the medium where 
one sees that for a gluon-induced shower its value is about ~32 GeV for our parameters.

We now evaluate the hydrodynamic response of the medium to a parton shower.  We consider the realistic scenario 
discussed directly above, only we now include both $\theta = 0.7$ and $\theta = 0.1$.  By comparing these two 
angles we can illustrate clearly the effect of the interference on the medium response.  The implementation 
of the full source term is rather numerically expensive therefore in this exploratory study we implement 
the simplest possible description of the shower by treating each parton as a $\delta$ function source with 
a weight given by the energy loss rate to the medium: $J^\nu = [dE_i/dt(t)] \, u_i^\nu\delta(\bald{r}_i - \bald{u}_i t)$, where the subscript $i$ refers to a summation over each parton in the shower.  This simplified form of the source correctly accounts for energy and momentum conservation and allows us to 
incorporate the energy loss rates evaluated above.  We make the stipulation that the contribution coming from 
the quantum interference is associated with the radiated gluon rather than the primary quark.

We focus on the energy density perturbation, $\delta \epsilon$, generated in the medium by the parton shower.  
Within linearized hydrodynamics this quantity can be written in momentum space as~\cite{Neufeld:2010tz}
\begin{eqnarray}\label{eps}
\delta\epsilon ({\mathbf r},t) = \int \frac{d^4 k}{(2\pi)^4} e^{- i k \cdot x} 
\frac{i k J_L(k)  + J^0(k)(i \omega -  \Gamma_s \bald{k}^2)}
{\omega^2 -  c_s^2 \bald{k}^2 + i \Gamma_s \omega \bald{k}^2}\, . 
\end{eqnarray}
In the above equation, $c_s$ denotes the speed of sound and $\Gamma_s = \frac{4 \eta }{3 s T}$ is the sound 
attenuation length with $\eta/s$ being the shear viscosity to entropy density ratio in the medium.  
Also, the source vector has been divided into transverse and longitudinal parts: 
${\mathbf J} = \hat{\mathbf k} J_L + {\mathbf J}_T$.  In addition to the parameters listed above we choose the 
proposed minimum bound for the shear viscosity to entropy density ratio $\eta/s = 0.08$~\cite{Kovtun:2004de}. 
 Although shear viscosity estimates obtained from phenomenological studies of hydrodynamic simulations of the 
bulk matter produced in heavy ion experiments~\cite{Romatschke:2007mq} suggest a value perhaps 2-3 times 
larger than $\eta/s = 0.08$, we have chosen the minimum bound because it is particularly well suited for 
plotting the important features of the medium response. 

In Figure~\ref{epsplots} we present the result for $\delta\epsilon$ for the scenario and parameters discussed
above.  We consider a primary quark of energy 75~GeV which emits 4 gluons of energy 4.5 GeV.  The gluon 
emissions are uniformly distributed over the primary quark's path length of 5 fm, at which point we plot 
the medium response.  For the purpose of illustration we have chosen the gluon emissions to alternate between 
the positive and negative $x$ directions, although in practice they are randomly distributed in the $x-y$ plane.  
Additionally, we have scaled the energy density distributions associated with each parton by a factor of 
$|x_-| = |\bald{r}-\bald{r}_0 - \bald{u}(t - t_0)|$ to compensate for the broadening of the Mach cones.  
Here, $\bald{r}_0$ and $t_0$ are the initial position and time of the particular parton.

In the left and right panels of Figure~\ref{epsplots} we show the result for $\theta = 0.1$ and $\theta = 0.7$, respectively. 
One can see that for the narrow emission angle the parton shower generates a well defined Mach cone similar 
to what one would expect from a single primary parton~\cite{Neufeld:2008fi}.  However, the interference effects 
from such small angle emissions suppress the energy transfer such that the total energy gained by the medium is 
only moderately enhanced compared to the single primary parton (about 1.7 times larger for our parameters).  
Contrast this to previous studies which ignored the interference suppression and found significant enhancement 
even for exactly collinear radiation~\cite{Neufeld:2009ep}.  For the larger emission angle there is no longer a 
well defined Mach cone but rather a superposition of several distinct perturbations in the medium.  The interference 
effects for this angle actually lead to an enhancement in the overall energy gained by the medium, even relative to 
no interference.  Compared to the single primary parton the energy gain is about 2.7 times larger. However, the price 
for this gain in energy is that there is no longer a well defined Mach cone structure.

\section{Conclusions}
\label{conclude}

In this Letter we evaluated the source term $J^\nu$ of energy and momentum deposited by a parton shower in the 
QGP in closed analytic form. Our result includes for the first time the quantum interference effects between the color 
currents associated with the primary parton and the radiated gluon. We demonstrated that the interference 
term that spoils the naive classical superposition~\cite{Neufeld:2009ep} of incoherent energy and momentum transfer rates persists for time scales relevant for heavy ion phenomenology.  We showed that this term vanishes in the $t\rightarrow \infty$ limit, as it must.

We presented two applications of the source term derived in this paper: the evaluation of the differential rate
of energy transfer by a parton shower to the medium and the hydrodynamic response of the medium to a parton shower.  
The former should not be confused with the rate of collisional energy loss~\cite{finitetime}
 of the parent parton itself, which remains constant and relatively small. We found that the rate of energy and and momentum transfer from 
the parton shower to the medium depends sensitively on the angle of emission and number of gluons emitted. For collinear 
bremsstrahlung, the medium cannot resolve the primary parton and the radiated gluon and there is pronounced coherent 
suppression in the magnitude of the multi-parton source term. For realistic large angles of emission~\cite{angle},  
we actually find an enhancement in its amplitude. Using the GLV formalism to describe 
the medium-induced parton shower~\cite{Vitev:2007ve}, including bremsstrahlung gluon multiplicities,
energies and emission angles,  we estimated that for LHC applications~\cite{lhcrad}
a 100~GeV parton can transmit more than 1/3 of its energy to the QGP. In the future, we plan to conduct 
more detailed numerical simulations that include realistic geometry and expansion of the 
medium and improved calculations of parton shower formation in the QGP~\cite{scet}. We note, however,
that our findings suggest that the rate of jet energy dissipation in the medium may significantly 
complicate the experimental/jet medium background separation at RHIC and at the LHC \cite{LHC,LHCIII}. 

We embeded the newly-derived source term in linearized hydrodynamics to study the response of the QGP  
to the in-medium parton shower. We found that for small gluon emission angles and minimal shear viscosity
to entropy density ratio $\eta/s = 0.08$~\cite{Kovtun:2004de} the parton shower generates a well 
defined Mach cone similar to what one would expect from a single primary parton or from a phenomenological 
incoherent collinear superposition of a parton with gluons~\cite{Neufeld:2009ep}. For realistic 
multiple gluon emission at large angles~\cite{angle} there is an enhancement in the rate of
energy transfer to the medium not only from the development of the parton shower but also from the 
interference effects discussed in this Letter. This, in turn, yields a larger perturbation of 
the energy density in the medium but there is no longer a well defined Mach cone in the QGP.   
Our result is intuitive and easy to understand: each parton in the developing shower gradually 
becomes a separate source of energy deposition and excitation of the strongly-interacting medium.
The overlapping perturbations in very different spatial directions wipe out any distinct Mach 
cone structure. Our finding is critical for heavy ion experiments and phenomenology. It suggests that 
schematic treatments of the parton shower provide an inaccurate description of shockwave 
phenomena in the QGP.  It also suggests that high $p_T$ triggers are not likely to have well 
defined away-side associated shockwave structures, since the number of medium-induced gluons grows 
with the parent parton energy~\cite{Vitev:2007ve}.

\bibliographystyle{model1-num-names}
\bibliography{<your-bib-database>}




\end{document}